\documentclass[conference]{IEEEtran}
\IEEEoverridecommandlockouts

\usepackage{etoolbox}
\makeatletter
\patchcmd{\maketitle}
  {\twocolumn[{%
    \IEEEquantizevspace{\@maketitle}[\IEEEquantizedisabletitlecmds]{0pt}[-\topskip]{\baselineskip}{\@IEEENORMtitlevspace}{\@IEEEMINtitlevspace}%
    \@IEEEaftertitletext}]}
  {\twocolumn[{%
    \vspace*{-0.3em} 
    \@maketitle\vspace*{0.3em}}]} 
  {}{}
\makeatother

\usepackage{cite}
\usepackage{amsmath,amssymb,amsfonts}
\usepackage{algorithmic}
\usepackage{graphicx}
\usepackage{textcomp}
\usepackage{xcolor}
\usepackage{url}
\usepackage{booktabs}
\usepackage{multirow}
\usepackage{tikz}
\usepackage{comment}
\usetikzlibrary{arrows.meta,positioning,fit,calc,backgrounds}
\usepackage{pgfplots}
\pgfplotsset{compat=1.18}
\def\BibTeX{{\rm B\kern-.05em{\sc i\kern-.025em b}\kern-.08em
    T\kern-.1667em\lower.7ex\hbox{E}\kern-.125emX}}

\definecolor{cNone}{HTML}{2C7BE5}
\definecolor{cLaya}{HTML}{E8692C}

\begin{document}
\bstctlcite{BSTcontrol}

\title{When Do Decision Models Help LLM-Driven Network Discovery?}

\author{
  \IEEEauthorblockN{
  Alperen Örsdemir\IEEEauthorrefmark{1}\IEEEauthorrefmark{2}\hspace{0.4cm}
  Rochelle Xenia Mendoza Santos\IEEEauthorrefmark{2}\hspace{0.4cm}Wen Shei Ong\IEEEauthorrefmark{2}\hspace{0.4cm}Matthew Caesar\IEEEauthorrefmark{3}\hspace{0.4cm}Ertem Esiner\IEEEauthorrefmark{2}}
  \vspace{0.2cm}
  \IEEEauthorblockA{
    \IEEEauthorrefmark{1}\textit{Koç University, Türkiye \textbar{}} Email: aorsdemir22@ku.edu.tr\\
    \IEEEauthorrefmark{2}\textit{Illinois Advanced Research Center at Singapore, Singapore }\\
    Email: \{rochelle.santos, e.esiner, wenshei.ong\}@iarcs-create.edu.sg\\
    \IEEEauthorrefmark{3}\textit{University of Illinois Urbana-Champaign, USA \textbar{}}
    Email: caesar@illinois.edu\\
}

  \vspace{-0.5em}
}

\maketitle

\begin{abstract}

Network discovery, the identification of live hosts and their exposed services, is the first phase of every penetration test, and it is increasingly automated by harnesses driven by large language models (LLMs). To mitigate hallucinated or unsafe commands, recent work proposes inserting fast, non-generative decision models to validate actions prior to execution. This paper quantifies whether such validation improves discovery across three network scales ($n \in \{10, 100, 1{,}000\}$) in a simulation calibrated on a 30-node containerized testbed. Our findings show that chunked batch dispatch recovers 92\% to 100\% of open ports in large-scale networks, whereas single-host dispatch collapses to under 6\% due to context saturation and scan latency. Furthermore, at the measured proposal quality of the local model, decision gating increases its port recall from 75.5\% to 89.0\% on small networks and intercepts costly decoy tarpits, but it imposes additional latency on frontier models during early host discovery.

\end{abstract}


\section{Introduction}

The increasing connectivity of enterprise systems and critical infrastructure has expanded the external attack surface, exposing internal assets to possible exploitation. In response, regulatory standards such as the Payment Card Industry Data Security Standard (PCI DSS)~\cite{pcidss40} have mandated routine Vulnerability Assessment and Penetration Testing (VAPT) across various industries. The efficacy of any VAPT engagement depends on its initial reconnaissance phase, where practitioners map out the network to identify active assets. Because downstream evaluations cannot consider hidden targets, host discovery must construct a complete inventory of active hosts and exposed network services. However, reconnaissance scales poorly across modern environments, where analysts must work on a vastly larger and more complex address space within rigid constraints on operational time and manpower.

Historically, host discovery has been automated deterministically with static scripts running network scanners~\cite{lyon2009nmap,durumeric2013zmap}. These approaches struggle with increasing scale and complexity because they cannot infer context from partial findings, prioritize targets, or dynamically alter their scanning trajectory when faced with atypical network topology. To address these shortcomings, literature has explored the use of LLMs to automate network discovery based on real-time evidence~\cite{deng2024pentestgpt,xu2024autoattacker}. Recent approaches rely on an autonomous agent harness to equip the LLM with network utilities, enabling the model to iteratively issue further commands and parse returned outputs~\cite{demacedo2026harness}.

Despite these developments, unconstrained LLMs pose significant operational risks during host discovery when deployed autonomously, as they can hallucinate non-existent targets, issue duplicate probes, or execute unsafe network commands. In response, recent works have placed calibrated decision models between the LLM and its network tools to validate its actions prior to execution~\cite{santos2026calibrated}. These specialized, non-generative decision layers such as Laya, evaluate structured queries about a proposed action, such as approve, reject, or redirect, in milliseconds with probabilities~\cite{convai2026laya}.

While these gating mechanisms promise immediate and effective validation, their practical utility under realistic constraints has not been empirically examined. This paper therefore investigates the performance trade-offs of an intermediate decision layer within an autonomous network discovery harness. Using a simulation calibrated on a 30-node testbed running the Laya decision model alongside a local LLM, we systematically evaluate how decision gating impacts port recall, loop latency, time-to-full-coverage, and decoy exposure under strict time allocations across scaling network topologies. This work’s primary contributions are:

\begin{enumerate}
\item A simulated study on autonomous network discovery evaluating an LLM-driven harness with and without an intermediate decision model.
\item A comparative performance analysis of the two configurations across different network scales, evaluating port recall, per-loop latency, time-to-full-coverage, and decoy timeouts within a 10-minute budget.
\item Evidence that batch scanning configurations govern discovery yield on large topologies, that intermediate decision models benefit smaller LLMs while delaying complete coverage for larger ones, and that short time budgets mask the model's operational value.
\end{enumerate}

\section{Related Work}

Penetration testing, as defined by NIST SP 800-115, is structured into four distinct phases: planning, discovery, attack and reporting~\cite{nist800115}. Within this framework, the discovery phase establishes the operational bounds of the assessment by enumerating reachable hosts, mapping open network ports, and fingerprinting underlying services and operating systems. This process has leveraged automation to accelerate discovery. Stateless mass scanners such as ZMap~\cite{durumeric2013zmap} can scan the whole public IPv4 address space in under 45 minutes from one machine by batching probes efficiently. To manage uncertainty during targeted exploration, some formalizations framed network reconnaissance as a partially observable Markov decision process (POMDP)~\cite{sarraute2012pomdp} or evaluated automated probing trajectories using reinforcement learning~\cite{schwartz2019autonomous}. However, these methods scale poorly in practice, as solving the underlying models becomes intractable over large address spaces and they require explicit state and action models of the target network.

Recent literature has utilized LLMs to autonomously conduct reconnaissance through a harness. In these architectures, the harness serves as the run-time layer, mediating tool access, managing context memory, and executing commands proposed by the generative model~\cite{demacedo2026harness,yao2023react}. Early implementations embedded LLMs within iterative plan-act-report loops to direct and summarize Nmap sweeps~\cite{moskal2023scriptkiddie}, or enabled the agent to issue raw bash commands whose outputs are fed back into the prompt context~\cite{happe2023pwnd}. However, monolithic prompts scale poorly with increasing network size and complexity, since accumulating scan logs saturates finite context windows and increases inference latency~\cite{liu2024lost}. Recent works therefore partition the penetration testing workflow into dedicated modules. PentestGPT utilizes discrete sub-modules dedicated to high-level reasoning, command generation, and output parsing~\cite{deng2024pentestgpt}. Subsequent frameworks, such as AutoAttacker~\cite{xu2024autoattacker}, VulnBot~\cite{kong2025vulnbot}, and PentestAgent~\cite{shen2025pentestagent}, have further distributed the workflow across collaborative multi-agent teams.

To prevent autonomous agents from issuing malformed or redundant network commands, the harness requires validation layers prior to tool execution. LLM-as-a-judge patterns deploy secondary generative models to evaluate proposed actions~\cite{zheng2023judging}, but introduce notable latency and computational overhead that hinders active scanning. Inspired by Kahneman’s dual-process architecture~\cite{kahneman2011thinking}, recent frameworks have paired fast, non-generative decision models (System One models) with slow, deliberate LLMs (System Two models). Rather than generating sequential tokens, these System One models evaluate structured state queries (multiple choice, binary, or numeric scoring) with calibrated probabilities in a single forward pass~\cite{typesafe2026jev}. Laya pairs a ModernBERT encoder backbone with a calibrated decision head~\cite{warner2024modernbert,convai2026laya}, yielding inference latencies of 33 to 40 ms per request on a T4 GPU~\cite{santos2026calibrated}. In hybrid decision-model and LLM frameworks, the decision model filters inbound queries and only escalates the query to an LLM judge when its confidence is below a certain threshold $\tau$. Such a cascade matches the accuracy of a full LLM judge at a fraction of its cost, with a median decision time of 0.27 s against 1.89 s~\cite{li2026jevjudge,rao2026decision}, although decision models remain sensitive to input framing and prompt perturbation~\cite{hu2026jevadvbench}.

How the per-loop overhead of decision models affects discovery across expanding networks remains unexamined. The closest work is by dos Santos~\cite{santos2026calibrated}, who proposed placing calibrated System One models, including Laya, across four decision points of an offensive security harness, and reported findings from an exploratory single-run case study. In contrast, this paper isolates the initial host discovery task and systematically evaluates the performance trade-offs of an intermediate decision layer under strict operational budgets. Furthermore, we scale the address space across three orders of magnitude and evaluate performance across 20 seeded runs per configuration, directly quantifying port recall, loop latency, time-to-full-coverage, and decoy timeouts.

\section{Autonomous Network Discovery Module}
\label{sec:module}

\begin{figure}[t]
\centering
\begin{tikzpicture}[font=\scriptsize, >={Stealth[length=1.6mm]},
  box/.style={draw, rounded corners=2pt, align=center, minimum height=6.5mm, minimum width=18mm, inner sep=2pt, fill=white},
  dm/.style={box, draw=cLaya, thick}]

\node[box] (kb)  at (0, 0)     {Knowledge Base};
\node[box] (pr)  at (2.8, 0)   {Prompt};
\node[box] (llm) at (5.6, 0)   {LLM};

\node[box] (jg)  at (2.8, -1.8) {LLM Judge};
\node[dm]  (dm)  at (5.6, -1.8) {Decision Model};

\coordinate (top-jg) at (2.8, -1.05);
\coordinate (top-dm) at (5.6, -1.05);

\begin{scope}[on background layer]
  \node[draw=cLaya, thick, dashed, fill=cLaya!8, rounded corners=4pt,
        fit=(jg) (dm) (top-jg) (top-dm),
        inner xsep=8pt, inner ysep=5pt] (layer) {};

  \node[font=\bfseries\scriptsize, text=cLaya, anchor=north] 
        at ([yshift=-4pt]layer.north) {Decision Layer};
\end{scope}

\node[box, minimum width=42mm] (sc) at (4.2, -3.5) {MCP Server (Nmap)};

\draw[->] (kb) -- (pr);
\draw[->] (pr) -- (llm);

\draw[->] (llm) -- node[pos=0.2, left]{proposals} (dm);

\draw[->] (dm.west) -- node[below, font=\tiny]{conf.\ $<\tau$} (jg.east);

\draw[->] (dm.south) -- node[pos=0.8, right]{approved} (dm.south |- sc.north);
\draw[->] (jg.south) -- node[pos=0.8, left]{approved} (jg.south |- sc.north);

\draw[->, dashed] (layer.north -| pr.south) -- node[pos=0.5, right]{rejected} (pr.south);

\draw[->] (sc.west) -| node[pos=0.85, left]{results} (kb.south);

\draw[->, dotted, thick, cNone] (llm.east) -- ++(3.5mm,0) |- node[pos=0.25, sloped, above, inner sep=1pt, text=cNone]{LLM only} (sc.east);

\end{tikzpicture}
\caption{The autonomous network discovery module with the decision layer.}
\label{fig:harness}
\vspace{-4mm} 
\end{figure}
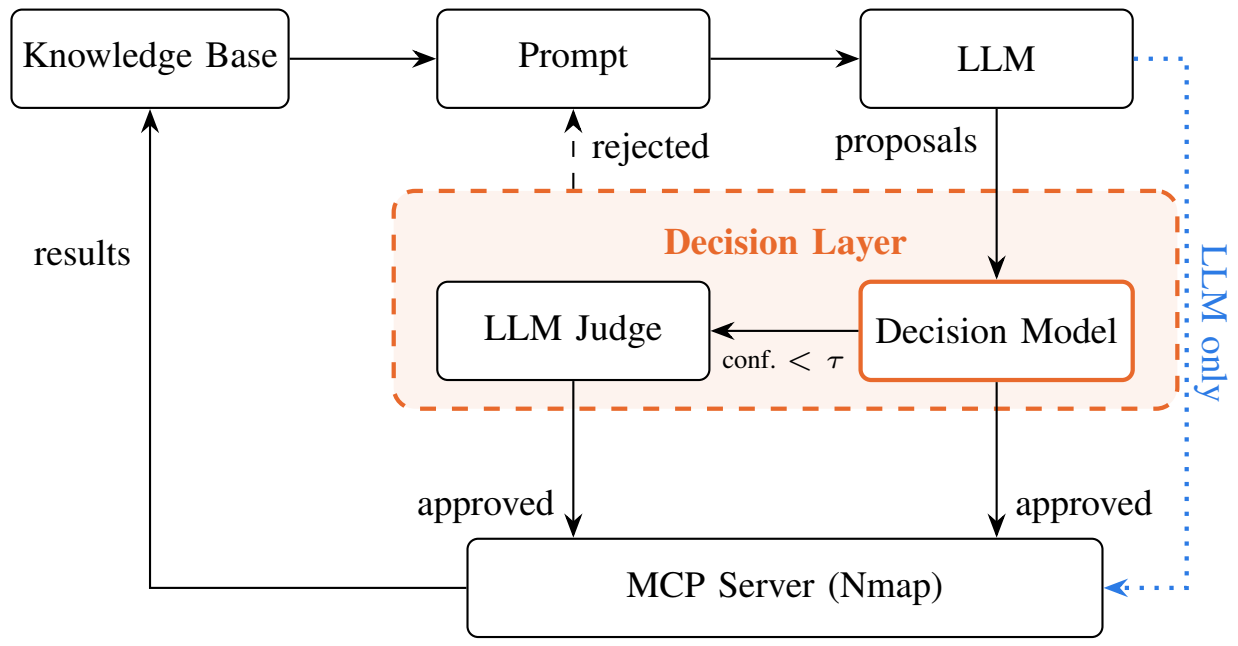

The autonomous network discovery module embeds an LLM within an execution harness consisting of a configurable decision layer with a calibrated decision model and an LLM judge, a knowledge base for accumulated information, and a Model Context Protocol (MCP) server that interfaces with the Nmap network scanner. It conducts reconnaissance in three distinct phases: host enumeration within a target CIDR scope, port discovery across identified live targets, and service and operating system fingerprinting. For each stage, the module executes an iterative orchestration loop (Fig.~\ref{fig:harness}). Each iteration begins by serializing information from the knowledge base into the LLM prompt context. The model proposes the next command, which is evaluated by the decision layer. Once approved, the harness dispatches the designated command to the Nmap scanner via the MCP server with an execution time limit. The resulting scan telemetry is parsed and used to update the knowledge base, which is then used to construct subsequent queries.

The module supports two execution configurations to evaluate the operational trade-offs of decision gating. In the \emph{LLM-only} baseline, the harness bypasses the decision layer entirely, automatically dispatching the LLM-proposed command to the MCP server. In contrast, the \emph{decision-gated} configuration prompts the LLM to propose $m=3$ candidate actions. The Laya decision model evaluates each candidate independently, computing a calibrated probability of execution validity. If the model confidence meets or exceeds a predefined operational threshold $\tau$, the first approved candidate command is dispatched to the MCP server. When confidence falls below $\tau$, the harness escalates evaluation to the LLM judge. If all proposed actions are rejected by the decision layer, the harness triggers a single reprompting round, appending the rejected candidates and associated error telemetry to the prompt so the model can rectify its proposal.

Furthermore, we consider two dispatch paradigms for delivering LLM-generated commands to the MCP server. In the \emph{per-host} loop, implemented in our harness, the LLM receives the current inventory of discovered hosts and open ports, and is prompted to propose a scan action targeting an individual host. Meanwhile, the \emph{chunked-plan} loop, which we evaluate in simulation only, prompts the model to propose an action across target hosts using an allocation chunk size. The harness partitions the target space into discrete slices and dispatches the scans sequentially without intermediate re-prompting. Under this configuration, subsequent prompts contain only an aggregated network summary and the targets within the active slice, avoiding context window bloat. When decision gating is enabled, the LLM proposes three candidate plans varying in chunk size, and each candidate is validated by the decision layer prior to dispatch.

\section{Simulation Methodology}

We evaluate the harness in a discrete-event simulation calibrated on a physical testbed: a Docker topology of 30 containerized nodes on isolated networks, on which the harness executes with a live Laya instance and a locally served Llama 3.1 8B. The testbed provides the measured scan durations of each host profile and the latency, context capacity and proposal quality of the local model (3 of 38 proposals useful, 19 of 38 repeated). The simulated target networks span three scales, with host counts $n \in \{10, 100, 1{,}000\}$, live-host open ports $P \in \{10, 136, 1{,}295\}$ and decoy counts $d \in \{1, 1, 10\}$, respectively. Each network combines the testbed host profiles with generated servers exposing one to four common services; 15\% of live hosts are workstations with no open ports. Decoys operate as tarpits that advertise 780 open ports and stall fingerprinting until the $300\text{ s}$ scanner timeout.

To evaluate the operational impact of intermediate validation, the simulation benchmarks the autonomous network discovery module running under the \emph{LLM-only} configuration and the \emph{decision-gated} configuration. Both configurations are evaluated using the \emph{per-host} and the \emph{chunked-plan} dispatch paradigms described in Section~\ref{sec:module}.

\subsection{Simulation Parameters and Metrics}
Each simulation trial operates under a budget of $B = 600\text{ s}$. The simulation clock advances cumulatively across each orchestration loop by aggregating generative inference latency $t^{\mathrm{llm}}$, decision validation overhead $t^{\mathrm{dm}}$, judge escalation duration $t^{\mathrm{j}}$, and network scanning runtime $t^{\mathrm{scan}}$. Across $K$ completed loops, total execution time ($T_K$) satisfies

\begin{equation}
T_K = \sum_{k=1}^{K} \left( t^{\mathrm{llm}}_k + t^{\mathrm{dm}}_k + t^{\mathrm{j}}_k + t^{\mathrm{scan}}_k \right) \le B.
\label{eq:clock}
\end{equation}

Any scanning operation active when the clock reaches $B$ is truncated and yields no telemetry. Individual host scanning durations $s(h)$ reflect completion times measured on the containerized testbed for its host types, whereas generated servers use assumed durations. When probing batched targets in parallel, total scan duration $t^{\mathrm{scan}}$ equals the slowest host duration plus an assumed per-host increment ($\delta_{\mathrm{survey}} = 0.15\text{ s}$ during host surveys and $\delta_{\mathrm{fp}} = 0.80\text{ s}$ during fingerprinting), bounded by a global scanner timeout of $T_{\mathrm{timeout}} = 300\text{ s}$.

System performance is evaluated across six primary metrics. We measure discovery yield through port recall ($R_P$), the proportion of live hosts' open ports identified within the time budget, and host recall ($R_H$), the proportion of target hosts identified. Operational overhead is characterized by mean loop duration ($\bar{t} = T_K / K$), total completed loops ($K$), and the elapsed duration required to reach complete port enumeration ($T_{100}$). Susceptibility to defensive deception is quantified through the decoy timeout count ($D$), denoting the number of trials in which fingerprinting sweeps were trapped by a tarpit for the full 300 s execution limit. To ensure statistical reproducibility, each experimental configuration is evaluated across 20 trials with fixed pseudo-random seeds, with comparative significance assessed using Welch's $t$-test, whose p-value is denoted $p$.

\subsection{Model and Harness Parameters}

Following Dhakal et al.~\cite{dhakal2026baselines}, who observe that performance deltas in agentic harnesses frequently reflect underlying model capabilities rather than framework architecture, we isolate the contribution of decision gating by evaluating two models under identical conditions: a locally served Llama 3.1 8B instance through Ollama~\cite{ollama} using a 4,096-token context window alongside an anonymous commercial hosted LLM representing a high-capacity frontier model, whose parameters are assumed.

Table~\ref{tab:params_llm} details the comparative parameters and proposal quality distributions across both generative models. Generative response latency incorporates a base reply duration $\alpha$ and a prompt-processing coefficient $\beta$, with an initial memory loading overhead $\lambda$ incurred on the first LLM invocation. Each response is useful with probability $u$, repeats completed work with probability $\rho$, and is invalid otherwise; under the chunked-plan loop, each model initially proposes a chunk size $C_1$. In the \emph{per-host} dispatch loop, accumulating target logs eventually saturate the local model context window ($W = 4{,}096$), triggering prompt truncation and degrading the useful proposal rate $u$ to a quarter of its baseline. Conversely, the \emph{chunked-plan} dispatch loop bounds batch allocations to $c \le c_{\max} = 162$ hosts for $W = 4{,}096$, ensuring prompt lengths remain within $W$, whereas the hosted frontier model operates with a non-binding context ceiling exceeding $128\text{k}$ tokens.

Operational thresholds and latency constants for the decision layer are cataloged in Table~\ref{tab:params_laya}. Candidate commands are filtered by the decision layer, which introduces an evaluation latency of $a + bq$ for $q$ questions per request. When classification confidence falls below operational threshold $\tau$, occurring with probability $\eta = 0.25$, evaluation escalates to the LLM judge, which resolves proposals with accuracy $\psi$. For independent classifications, the decision model intercepts redundant commands, malformed actions, and known decoys with recall probabilities $\pi_{\mathrm{rep}}$, $\pi_{\mathrm{bad}}$, and $\pi_{\mathrm{dec}}$, while exhibiting a false rejection rate $\pi_{\mathrm{fr}}$ against valid actions.

\begin{table}[t]
\centering
\scriptsize
\setlength{\tabcolsep}{4.5pt}
\setlength{\belowcaptionskip}{2pt}
\caption{LLM Agent Calibration and Simulation Parameters}
\label{tab:params_llm}
\begin{tabular}{@{}llcc@{}}
\toprule
\textbf{Parameter} & \textbf{Description} & \textbf{Llama 3.1 8B} & \textbf{Frontier LLM} \\
& & \textit{(Testbed)} & \textit{(Assumed)} \\
\midrule
$\alpha$ (s) & Mean response latency & 2.5 \textit{(M)} & 6.0 \\
$\beta$ (s) & Latency per 1k prompt tokens & 1.5 & 0.2 \\
$\lambda$ (s) & First-call cold-start load & 8.9 \textit{(M)} & 0.0 \\
$W$ (tokens) & Context window capacity & 4,096 \textit{(M)} & $\ge$128k$^{\dagger}$ \\
$u$ & Useful response share & 0.08 \textit{(M)} & 0.85 \\
$\rho$ & Repeat query share & 0.50 \textit{(M)} & 0.10 \\
$\psi$ & Evaluator / judge accuracy & 0.70 & 0.90 \\
$C_1$ (hosts) & Initial batch chunk size & 25 & 64 \\
\bottomrule
\multicolumn{4}{@{}l}{\textit{(M)}: Empirically measured on testbed; all others are assumed.} \\
\multicolumn{4}{@{}l}{$^{\dagger}$Never binding: maximum benchmark prompt is 16.2k tokens.}
\end{tabular}
\vspace{-2mm} 
\end{table}

\begin{table}[t]
\centering
\scriptsize
\setlength{\tabcolsep}{6pt}
\setlength{\belowcaptionskip}{2pt}
\caption{Decision Model (Laya) Parameters}
\label{tab:params_laya}
\begin{tabular}{@{}llr@{}}
\toprule
\textbf{Parameter} & \textbf{Description} & \textbf{Assumed Value} \\
\midrule
$a$ & Latency overhead per request & 0.30 s \\
$b$ & Incremental overhead per question & 0.03 s \\
$\eta$ & Traffic fraction below score threshold $\tau$ & 0.25 \\
$\pi_{\mathrm{rep}}$ & Rejection accuracy for repeated queries & 0.85 \\
$\pi_{\mathrm{bad}}$ & Rejection accuracy for malformed / invalid queries & 0.90 \\
$\pi_{\mathrm{dec}}$ & Rejection accuracy for known decoy targets & 0.90 \\
$\pi_{\mathrm{fr}}$ & False rejection rate for viable targets & 0.05 \\
\bottomrule
\end{tabular}
\vspace{-2mm} 
\end{table}

\section{Experimental Results}

Table~\ref{tab:results} summarizes experimental performance across 24 configurations evaluated under the $B = 600\text{ s}$ operational budget, covering the three network scales, two dispatch paradigms, two generative models, and the presence or absence of decision gating. Across all evaluated conditions, every configuration achieves complete host recall ($R_H = 1.0$), as initial ping sweeps reliably enumerate reachable addresses within the defined CIDR scope. This section examines the performance of the \emph{LLM-only} and \emph{decision-gated} network discovery modules, compares the decision-gated performance of two generative models of different sizes, and analyzes the sensitivity of the results to the decision model parameters.

\begin{table}[t]
\caption{Performance of LLM-Only and Decision-Gated Network Discovery}
\label{tab:results}
\centering
\scriptsize
\setlength{\tabcolsep}{2.6pt}
\begin{tabular}{@{}lll rrrr rrrr@{}}
\toprule
 & & & \multicolumn{4}{c}{LLM only} & \multicolumn{4}{c}{LLM + Laya}\\
\cmidrule(lr){4-7}\cmidrule(l){8-11}
$n$ & LLM & Loop & $R_P$ & $\bar t$ & $K$ & $D$ & $R_P$ & $\bar t$ & $K$ & $D$\\
\midrule
\multirow{4}{*}{10} & Llama & per-host & 75.5 & 5.2 & 117 & 0 & 89.0 & 12.4 & 48 & 0 \\
 & Llama & chunked & 100.0 & 13.9 & 43 & 12 & 100.0 & 16.1 & 15 & 2 \\
 & frontier & per-host & 100.0 & 23.4 & 26 & 15 & 100.0 & 21.4 & 28 & 0 \\
 & frontier & chunked & 100.0 & 71.0 & 8 & 20 & 100.0 & 36.9 & 5 & 5 \\
\midrule
\multirow{4}{*}{100} & Llama & per-host & 5.9 & 6.3 & 95 & 0 & 9.9 & 12.5 & 48 & 0 \\
 & Llama & chunked & 100.0 & 12.4 & 48 & 5 & 100.0 & 17.4 & 27 & 1 \\
 & frontier & per-host & 52.2 & 8.3 & 73 & 0 & 47.6 & 10.7 & 56 & 0 \\
 & frontier & chunked & 100.0 & 56.6 & 11 & 20 & 100.0 & 31.9 & 8 & 1 \\
\midrule
\multirow{4}{*}{1,000} & Llama & per-host & 0.0 & 12.3 & 49 & 0 & 0.2 & 18.9 & 32 & 0 \\
 & Llama & chunked & 91.6 & 10.1 & 59 & 0 & 83.3 & 12.0 & 50 & 0 \\
 & frontier & per-host & 5.4 & 10.8 & 56 & 0 & 4.9 & 13.4 & 45 & 0 \\
 & frontier & chunked & 100.0 & 30.9 & 19 & 0 & 99.0 & 25.9 & 23 & 0 \\
\bottomrule
\multicolumn{11}{@{}l}{\textit{Models:} Llama = Llama 3.1 8B (local); frontier = commercial hosted LLM.} \\ 
\multicolumn{11}{@{}l}{\textit{Metrics:} $R_P$: port recall (\%); $\bar t$: mean loop time (s); $K$: loops completed;}  \\
\multicolumn{11}{@{}l}{$D$: decoy timeouts (max 20).} \\
\end{tabular}
\vspace{-2mm} 
\end{table}

\begin{figure}[t]
\centering
\begin{tikzpicture}
\begin{axis}[
  width=\linewidth,
  height=4.2cm,
  xmin=0, xmax=600,
  ymin=0, ymax=102,
  xtick={0,60,120,180,240,300,360,420,480,540,600},
  xticklabels={0,1,2,3,4,5,6,7,8,9,10},
  xlabel={Time (min)},
  ylabel={Port recall $R_P(t)$ (\%)},
  label style={font=\scriptsize},
  tick label style={font=\scriptsize},
  ymajorgrids,
  grid style={gray!20},
  axis line style={gray!60},
  tick style={gray!60},
  legend to name=namedlegend:cov,
  legend style={
    font=\scriptsize,
    draw=none,
    fill=none,
    legend columns=2,
    /tikz/every even column/.append style={column sep=3mm}
  },
  every axis plot/.append style={line width=1pt, no markers}
]
\addplot[cNone] table[x=t, y=r]{data/cov_large_none.dat};
\addplot[cNone, dashed] table[x=t, y=r]{data/cov_llama_none.dat};
\addplot[cLaya] table[x=t, y=r]{data/cov_large_laya.dat};
\addplot[cLaya, dashed] table[x=t, y=r]{data/cov_llama_laya.dat};
\legend{Frontier LLM only, Llama3.1 only, Frontier LLM + Laya, Llama3.1 + Laya}
\end{axis}
\end{tikzpicture}
\vspace{-1mm}
\ref{namedlegend:cov}
\vspace{-2mm}
\caption{Port recall over time at 1,000 hosts with the chunked-plan loop (mean of 20 runs).}
\label{fig:cov}
\vspace{-4mm}
\end{figure}
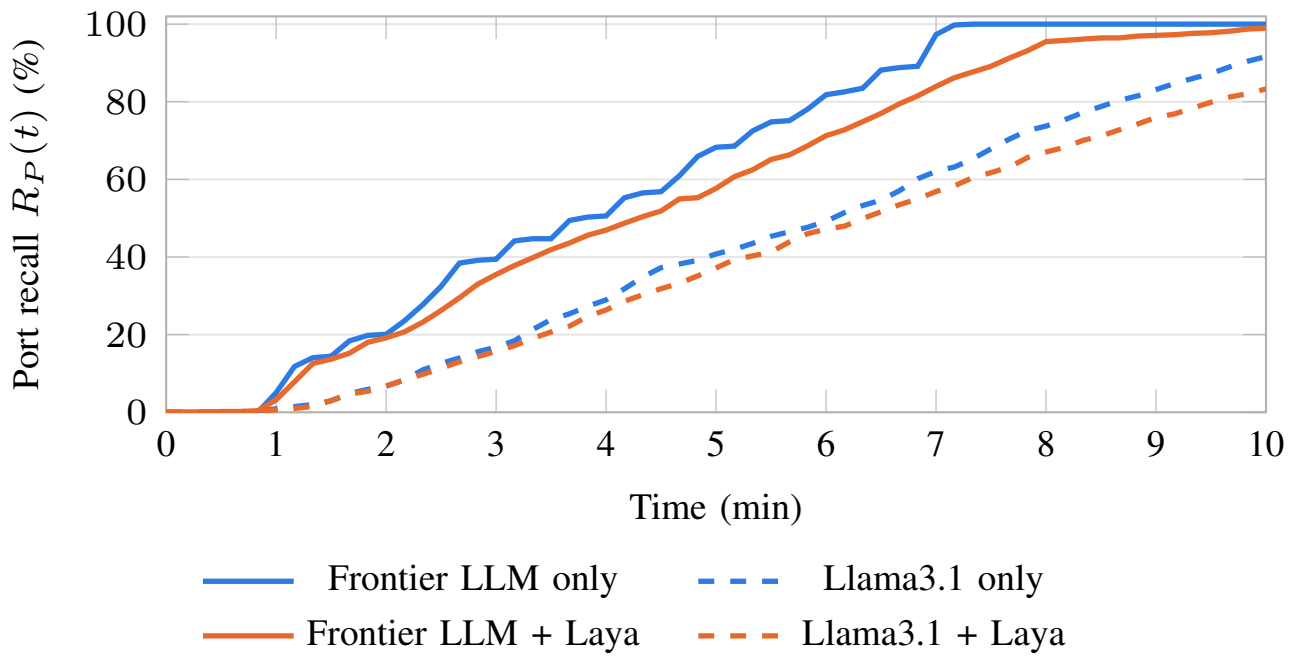

\subsection{LLM-Only Network Discovery}

Without intermediate decision gating, \emph{per-host} dispatch degrades sharply as network scale increases. The frontier model achieves complete port recall ($R_P = 100.0\%$) only at $n = 10$, falling to $52.2\%$ at $n = 100$ and $5.4\%$ at $n = 1{,}000$. Degradation is more severe for Llama 3.1 8B, which drops from $75.5\%$ recall at $n = 10$ to $5.9\%$ at $n = 100$, and fails entirely ($R_P = 0.0\%$) at $n = 1{,}000$. This failure stems from two compounding bottlenecks. First, issuing queries for individual targets incurs severe cumulative scan latency. With mean iteration times between 11 s and 12 s at scale, the operational budget permits only approximately 50 iterations. Even with an idealized stub planner replacing LLM reasoning, single-host sweeps across 1,000 targets require 1,713 s of aggregate scan time, exhausting the budget while surveying only 387 hosts and recovering 508 of 1,295 ports. Second, accumulating scan logs saturates context memory. At $n = 1{,}000$, knowledge base logs expand prompt lengths toward 16,000 tokens, far exceeding local model limits. Furthermore, at $n = 100$, encountering a single high-port decoy expands subsequent prompts from 2,248 to 4,812 tokens, triggering truncation that reduces proposal utility.

Conversely, \emph{chunked-plan} dispatch resolves both limitations. Because bounded batch allocation keeps prompt sizes within context limits, chunked dispatch achieves $R_P = 100.0\%$ across both models at $n \in \{10, 100\}$. At $n = 1{,}000$, chunking recovers $91.6\%$ of ports for Llama 3.1 8B and $100.0\%$ for the frontier model, outperforming per-host dispatch by 91.6 and 94.6 percentage points, respectively.

\subsection{Decision-Gated Network Discovery}

As shown in Table~\ref{tab:results}, decision gating improves port recall for Llama 3.1 8B under per-host dispatch on small and medium topologies, increasing recall from 75.5\% to 89.0\% at $n = 10$ and from 5.9\% to 9.9\% at $n = 100$. However, the decision model does not yield better performance under chunked dispatch at scale ($n = 1{,}000$), where port recall decreases by 8.3 points for Llama 3.1 8B ($p = 0.008$) and by 1.0 point for the frontier model, which is not significant ($p = 0.11$). Fig.~\ref{fig:cov} shows that gated and un-gated configurations track closely during initial exploration, but diverge as decision gating consumes 9\% to 14\% of the total execution budget on validation and judge escalation. 

This latency overhead is driven primarily by judge escalation rather than the decision model itself. Across loops at $n = 1{,}000$, the Laya decision model requires only 0.4~s to 0.6~s per iteration, whereas escalating uncertain proposals to the LLM judge adds 1.2~s to 3.1~s. Under \emph{per-host} dispatch with Llama 3.1 8B, decision gating increases mean loop duration from 12.3~s to 18.9~s, of which judge escalation accounts for 3.1~s. Conversely, under \emph{chunked-plan} dispatch, network scanning remains the primary operational bottleneck, accounting for 8.6~s to 29.7~s of total loop time across both models.

The primary operational benefit of intermediate decision gating lies in decoy mitigation during service fingerprinting. In un-gated configurations on small and medium topologies ($n \in \{10, 100\}$), fingerprinting sweeps frequently target high-port decoys, incurring the full 300~s execution timeout in up to 20 of 20 trials and consuming half of the experimental budget. Incorporating Laya intercepts these sweeps, reducing decoy timeouts to at most five occurrences. At $n = 1{,}000$, the budget expires during host enumeration or early fingerprinting, before any decoy is reached, resulting in zero decoy timeouts ($D = 0$) across all trials.

\subsection{Effect of Decision Gating Across Model Scales}

The operational value of intermediate decision gating depends heavily on the generative capability of the underlying model. For Llama 3.1 8B, the \emph{per-host} gains at $n \in \{10, 100\}$, which raise the trials reaching complete discovery at $n = 10$ from 4 to 12 of 20, stem from multi-candidate selection. Because Llama 3.1 8B has a low baseline proposal utility ($u = 0.08$), evaluating $m = 3$ candidates yields a 2.8-fold gain in actionable commands (to 0.22), which easily offsets the additional 7~s per loop incurred by validation and reprompting. As $u$ is estimated from only 3 of 38 testbed proposals, these gains should be read as indicative of the measured proposal quality rather than as precise estimates.

Conversely, high-capacity frontier models experience diminishing returns from candidate filtering. Because the frontier model already exhibits high baseline utility ($u = 0.85$), evaluating three candidates increases the viable proposal rate by only a factor of 1.17. At $n = 10$, where both configurations reach complete port recall, decision gating yields no recall benefit and delays complete coverage from 126~s to 135~s under \emph{per-host} dispatch and from 39~s to 44~s under chunked dispatch, with decoy avoidance as its only gain. At $n = 100$ under \emph{per-host} dispatch, this latency overhead causes port recall to drop from 52.2\% to 47.6\%. Thus, intermediate decision gating provides substantial improvement for smaller, error-prone models, whereas for frontier models, whose un-gated proposals are already highly reliable, it mainly delays complete coverage and its benefit is limited to decoy avoidance.

\subsection{Decision Model Sensitivity}
\label{sec:sens}

We further evaluate the impact of the decision layer on performance by varying the parameters of the decision model (Table~\ref{tab:sens}). Under chunked dispatch at scale ($n = 1{,}000$), an idealized profile featuring zero classification error, zero escalation ($\eta = 0$), and minimal latency (0.04~s) yields at most a 3.0-point gain in port recall, which is not significant, for Llama 3.1 8B while leaving frontier model performance unchanged. Across all parameter sweeps, judge escalation ($\eta$) governs operational overhead. Escalating half of proposed actions to the LLM judge ($\eta = 0.50$) consumes 23.4\% of the execution budget and degrades port recall by 19.3 points, whereas eliminating escalation entirely limits overhead to 3.1\%. Accelerating decision model inference alone recovers only 2.0 points of port recall, confirming that judge escalation latency rather than classification speed constitutes the primary operational bottleneck.

\begin{table}[t]
\caption{Sensitivity: Change in Port Recall With Laya (Points) and Decision Overhead (\% of Budget)}
\label{tab:sens}
\centering
\scriptsize
\setlength{\tabcolsep}{3pt}
\begin{tabular}{@{}lrrrr@{}}
\toprule
 & \multicolumn{2}{c}{1,000 hosts, chunked} & 10, per-host & Overhead\\
\cmidrule(lr){2-3}\cmidrule(lr){4-4}\cmidrule(l){5-5}
Laya profile & Llama3.1 & Frontier & Llama3.1 & Llama3.1\\
\midrule
Default (Table~\ref{tab:params_laya}) & $-$8.3$^{*}$ & $-$1.0 & +13.5$^{*}$ & 13.6 \\
Reported latency, 0.04 s & $-$6.3$^{*}$ & $-$0.9 & +14.0$^{*}$ & 11.1 \\
No escalation, $\eta=0$ & +2.4 & 0.0 & +21.5$^{*}$ & 3.1 \\
High escalation, $\eta=0.5$ & $-$19.3$^{*}$ & $-$2.4 & +3.5 & 23.4 \\
More false rejects, $\pi_{\mathrm{fr}}=0.15$ & $-$12.3$^{*}$ & $-$1.0 & +11.5$^{*}$ & 15.1 \\
Ideal & +3.0 & 0.0 & +24.5$^{*}$ & 0.4 \\
\bottomrule
\multicolumn{5}{@{}l}{$^{*}p<0.05$, Welch’s $t$-test against LLM only (20 runs each).}
\end{tabular}
\vspace{-2mm}
\end{table}

\section{Simulation Scope and Limitations}

While our simulation incorporates heterogeneous host profiles, non-responsive workstations and decoy tarpits, its results also highlight constraints that motivate future extensions.

At scale ($n = 1{,}000$), the gated and un-gated configurations differed by at most 8.3 percentage points in port recall, underscoring the trade-off between the decision layer overhead and averted scan penalties. Decision gating pays off only when the time it saves exceeds its overhead, and this saving varies by reconnaissance stage. During host discovery, inefficient commands carry negligible execution penalties but incur substantial validation overhead. In contrast, intercepting a single decoy scan in service fingerprinting averts a 300~s timeout penalty. The design is therefore compatible with stage-selective gating, routing only costly actions such as fingerprinting sweeps through the decision layer to retain decoy interception without discovery-phase overhead, although small LLMs would then forgo their gain from multi-candidate selection. Because enumerating hosts over a large address space consumes considerable time, most configurations exhausted the uniform 600~s budget before reaching service fingerprinting, where the benefits of decision gating would be more pronounced. Future evaluations should thus scale the budget proportionally to network size, as practical penetration testing would allocate more time to larger networks, and should compare decision gating against a rule-based filter, such as skipping repeated targets and hosts advertising hundreds of open ports, which would likely intercept the tarpit decoys modeled here.

The immediate next step is evaluating the performance of the network discovery module in a practical network environment. While the simulation captures the empirical baselines in a controlled setting, physical deployments introduce additional factors that can compound scan durations such as packet drops, buffer queueing, and asymmetric routing. Moreover, while local model latency can be reliably profiled, that of commercial models fluctuates, so live evaluations should track the latency of all LLM-driven tasks dynamically instead of relying on parametric assumptions.

Aside from live network evaluation, future work can focus on expanding the module’s capabilities and resilience. First, multi-port host validation can counter deceptive firewalls and honeypots that forge responses for unallocated addresses, by verifying candidate hosts before deeper fingerprinting. Second, the module can be extended to operational technology protocols such as MQTT, Modbus, and CoAP, where aggressive scanning risks disrupting critical hardware~\cite{nist80082r3}; tuning the decision model to recognize industrial assets and veto such probes would protect availability during discovery. Finally, because scan output is controlled by the target, a hostile host can forge telemetry, such as fake open ports or service banners, that misleads both the LLM and the decision model~\cite{hu2026jevadvbench}; our simulation assumes honest scan output, so this threat remains to be quantified.

\section{Conclusion}

This paper systematically evaluated the effect of an intermediate decision layer within an autonomous network discovery harness. Across simulated network scales and dispatch paradigms, our results demonstrate that the \emph{chunked-plan} dispatch loop resolves context memory saturation and per-host scan latency, enabling both local and frontier models to achieve high discovery yield at scale. Furthermore, at the measured proposal quality of the local model, intermediate decision gating improves discovery for smaller, error-prone models on small networks, yields no coverage gain at scale, and delays complete coverage for frontier models. Future work will focus on practical network evaluation, multi-port host validation, and tuning the decision model for sensitive operational technology environments.

\section*{Acknowledgment}
This research is supported by the National Research Foundation, Prime Minister’s Office, Singapore under its Campus for Research Excellence and Technological Enterprise (CREATE) programme.

\bibliographystyle{IEEEtran}
\bibliography{ref}

@IEEEtranBSTCTL{BSTcontrol,
  CTLuse_forced_etal        = "yes",   
  CTLmax_names_forced_etal  = "1",     
  CTLnames_show_etal        = "1",
  CTLdash_repeated_names   = "no", 
}

@techreport{nist800115,
  author      = {Scarfone, Karen and Souppaya, Murugiah and Cody, Amanda and Orebaugh, Angela},
  title       = {Technical Guide to Information Security Testing and Assessment},
  institution = {Nat. Inst. Standards Technol.},
  address     = {Gaithersburg, MD, USA},
  type        = {NIST Special Publication},
  number      = {800-115},
  month       = sep,
  year        = {2008},
  doi         = {10.6028/NIST.SP.800-115}
}

@inproceedings{deng2024pentestgpt,
  author    = {Deng, Gelei and Liu, Yi and Mayoral-Vilches, V{\'\i}ctor and Liu, Peng and Li, Yuekang and Xu, Yuan and Zhang, Tianwei and Liu, Yang and Pinzger, Martin and Rass, Stefan},
  title     = {{PentestGPT}: Evaluating and Harnessing Large Language Models for Automated Penetration Testing},
  booktitle = {33rd USENIX Security},
  address   = {Philadelphia, PA, USA},
  month     = aug,
  year      = {2024},
  pages     = {847--864}
}

@inproceedings{happe2023pwnd,
  author    = {Happe, Andreas and Cito, J{\"u}rgen},
  title     = {Getting pwn'd by {AI}: Penetration Testing with Large Language Models},
  booktitle = {31st ACM ESEC/FSE},
  address   = {San Francisco, CA, USA},
  month     = dec,
  year      = {2023},
  pages     = {2082--2086},
  doi       = {10.1145/3611643.3613083}
}

@misc{xu2024autoattacker,
  author       = {Xu, Jiacen and Stokes, Jack W. and McDonald, Geoff and Bai, Xuesong and Marshall, David and Wang, Siyue and Swaminathan, Adith and Li, Zhou},
  title        = {{AutoAttacker}: A Large Language Model Guided System to Implement Automatic Cyber-attacks},
  howpublished = {arXiv:2403.01038},
  month        = mar,
  year         = {2024}
}

@misc{kong2025vulnbot,
  author       = {Kong, He and Hu, Die and Ge, Jingguo and Li, Liangxiong and Li, Tong and Wu, Bingzhen},
  title        = {{VulnBot}: Autonomous Penetration Testing for a Multi-Agent Collaborative Framework},
  howpublished = {arXiv:2501.13411},
  month        = jan,
  year         = {2025}
}

@inproceedings{shen2025pentestagent,
  author    = {Shen, Xiangmin and Wang, Lingzhi and Li, Zhenyuan and Chen, Yan and Zhao, Wencheng and Sun, Dawei and Wang, Jiashui and Ruan, Wei},
  title     = {{PentestAgent}: Incorporating {LLM} Agents to Automated Penetration Testing},
  booktitle = {20th ACM ASIA CCS},
  address   = {Hanoi, Vietnam},
  month     = aug,
  year      = {2025},
  pages     = {375--391},
  doi       = {10.1145/3708821.3733882}
}

@misc{dhakal2026baselines,
  author       = {Dhakal, Ananda and Neupane, Krish and Chaudhary, Aarjan},
  title        = {Baselines Before Architecture: Evaluating Coding Agents for Autonomous Penetration Testing},
  howpublished = {arXiv:2607.13085},
  month        = jul,
  year         = {2026}
}

@misc{santos2026calibrated,
  author       = {dos Santos, Joas Antonio},
  title        = {Calibrated Decision Models for Autonomous Penetration-Testing Harnesses: {JEV} and {Laya} as {System One} Decision Layers for {LLM}-Driven Pentest Agents},
  howpublished = {arXiv:2609.28940},
  month        = sep,
  year         = {2026}
}

@inproceedings{yao2023react,
  author    = {Yao, Shunyu and Zhao, Jeffrey and Yu, Dian and Du, Nan and Shafran, Izhak and Narasimhan, Karthik and Cao, Yuan},
  title     = {{ReAct}: Synergizing Reasoning and Acting in Language Models},
  booktitle = {11th ICLR},
  address   = {Kigali, Rwanda},
  month     = may,
  year      = {2023}
}

@article{liu2024lost,
  author  = {Liu, Nelson F. and Lin, Kevin and Hewitt, John and Paranjape, Ashwin and Bevilacqua, Michele and Petroni, Fabio and Liang, Percy},
  title   = {Lost in the Middle: How Language Models Use Long Contexts},
  journal = {Trans. Assoc. Comput. Linguistics},
  volume  = {12},
  pages   = {157--173},
  year    = {2024},
  doi     = {10.1162/tacl_a_00638}
}

@book{kahneman2011thinking,
  author    = {Kahneman, Daniel},
  title     = {Thinking, Fast and Slow},
  publisher = {Farrar, Straus and Giroux},
  address   = {New York, NY, USA},
  year      = {2011}
}

@misc{typesafe2026jev,
  author       = {Almeida, Diogo},
  title        = {Introducing {System One} Models \& {Jev}},
  howpublished = {TypeSafe AI Blog},
  month        = sep,
  year         = {2026},
  url          = {https://typesafe.ai/blog/introducing-system-one-models-and-jev},
  note         = {Accessed: Oct. 26, 2026}
}

@misc{convai2026laya,
  author       = {{Convai Innovations}},
  title        = {{Laya}: Multilingual, Non-Autoregressive {System 1} Decision Model},
  howpublished = {Hugging Face model card},
  month        = sep,
  year         = {2026},
  url          = {https://huggingface.co/convaiinnovations/laya},
  note         = {Accessed: Oct. 26, 2026}
}

@inproceedings{warner2024modernbert,
  author    = {Warner, Benjamin and Chaffin, Antoine and Clavi{\'e}, Benjamin and Weller, Orion and Hallstr{\"o}m, Oskar and Taghadouini, Said and Gallagher, Alexis and Biswas, Raja and Ladhak, Faisal and Aarsen, Tom and Adams, Griffin Thomas and Howard, Jeremy and Poli, Iacopo},
  title     = {Smarter, Better, Faster, Longer: A Modern Bidirectional Encoder for Fast, Memory Efficient, and Long Context Finetuning and Inference},
  booktitle = {63rd ACL},
  address   = {Vienna, Austria},
  month     = jul,
  year      = {2025},
  pages     = {2526--2547},
  doi       = {10.18653/v1/2025.acl-long.127}
}

@inproceedings{zheng2023judging,
  author    = {Zheng, Lianmin and Chiang, Wei-Lin and Sheng, Ying and Zhuang, Siyuan and Wu, Zhanghao and Zhuang, Yonghao and Lin, Zi and Li, Zhuohan and Li, Dacheng and Xing, Eric P. and Zhang, Hao and Gonzalez, Joseph E. and Stoica, Ion},
  title     = {Judging {LLM}-as-a-Judge with {MT-Bench} and {Chatbot Arena}},
  booktitle = {Proc. NeurIPS, Datasets Benchmarks Track},
  address   = {New Orleans, LA, USA},
  volume    = {36},
  month     = dec,
  year      = {2023},
  pages     = {46595--46623}
}

@misc{li2026jevjudge,
  author       = {Li, Yubo and Miao, Yidi and Krishnan, Ramayya and Padman, Rema},
  title        = {{JEV}-as-a-Judge: Accept When Confident, Escalate When Unsure},
  howpublished = {arXiv:2609.26550},
  month        = sep,
  year         = {2026}
}

@misc{rao2026decision,
  author       = {Rao, Delip and Callison-Burch, Chris},
  title        = {Decision Models vs. {LLMs} as Rubric Judges: Cheaper, Faster, and Wrong in the Same Places},
  howpublished = {arXiv:2609.29769},
  month        = sep,
  year         = {2026}
}

@misc{hu2026jevadvbench,
  author       = {Hu, Jianyi and Zhang, Hangtao and Liu, Yi and Zeng, Yeqi and Zeng, Li and Wang, Xianlong and Wang, Rui and Zhang, Leo Yu},
  title        = {{JevAdvBench}: A Benchmark and Black-Box Attacks for Reinforcement Learning for Calibrated Decisions Models},
  howpublished = {arXiv:2609.31142},
  month        = sep,
  year         = {2026}
}

@book{lyon2009nmap,
  author    = {Lyon, Gordon Fyodor},
  title     = {Nmap Network Scanning: The Official {Nmap} Project Guide to Network Discovery and Security Scanning},
  publisher = {Insecure.Com LLC},
  address   = {Sunnyvale, CA, USA},
  year      = {2009}
}

@misc{pcidss40,
  author       = {{PCI Security Standards Council}},
  title        = {Payment Card Industry Data Security Standard: Requirements and Testing Procedures, Version 4.0},
  howpublished = {PCI SSC},
  month        = mar,
  year         = {2022}
}

@inproceedings{durumeric2013zmap,
  author    = {Durumeric, Zakir and Wustrow, Eric and Halderman, J. Alex},
  title     = {{ZMap}: Fast Internet-wide Scanning and Its Security Applications},
  booktitle = {22nd USENIX Security},
  address   = {Washington, DC, USA},
  month     = aug,
  year      = {2013},
  pages     = {605--620}
}

@inproceedings{sarraute2012pomdp,
  author    = {Sarraute, Carlos and Buffet, Olivier and Hoffmann, J{\"o}rg},
  title     = {{POMDPs} Make Better Hackers: Accounting for Uncertainty in Penetration Testing},
  booktitle = {26th AAAI},
  address   = {Toronto, ON, Canada},
  month     = jul,
  year      = {2012}
}

@misc{schwartz2019autonomous,
  author       = {Schwartz, Jonathon and Kurniawati, Hanna},
  title        = {Autonomous Penetration Testing Using Reinforcement Learning},
  howpublished = {arXiv:1905.05965},
  month        = may,
  year         = {2019}
}

@misc{moskal2023scriptkiddie,
  author       = {Moskal, Stephen and Laney, Sam and Hemberg, Erik and O'Reilly, Una-May},
  title        = {{LLMs} Killed the Script Kiddie: How Agents Supported by Large Language Models Change the Landscape of Network Threat Testing},
  howpublished = {arXiv:2310.06936},
  month        = oct,
  year         = {2023}
}

@misc{demacedo2026harness,
  author       = {de Macedo, Sanderson Oliveira},
  title        = {What Makes a Harness a Harness: Necessary and Sufficient Conditions for an Agent Harness},
  howpublished = {arXiv:2606.10106},
  month        = jun,
  year         = {2026}
}

@misc{ollama,
  author       = {{Ollama}},
  title        = {Ollama},
  howpublished = {GitHub repository},
  year         = {2026},
  url          = {https://github.com/ollama/ollama},
  note         = {Accessed: Oct. 26, 2026}
}

@techreport{nist80082r3,
  author      = {Stouffer, Keith and Pease, Michael and Tang, CheeYee and Zimmerman, Timothy and Pillitteri, Victoria and Lightman, Suzanne and Hahn, Adam and Saravia, Stephanie and Sherule, Aslam and Thompson, Michael},
  title       = {Guide to Operational Technology ({OT}) Security},
  institution = {Nat. Inst. Standards Technol.},
  address     = {Gaithersburg, MD, USA},
  type        = {NIST Special Publication},
  number      = {800-82r3},
  month       = sep,
  year        = {2023},
  doi         = {10.6028/NIST.SP.800-82r3}
}

\end{document}